\documentclass[prd,preprintnumbers,nofootinbib,tightenlines,superscriptaddress]{revtex4}
\pdfoutput=1
\usepackage{graphicx,epstopdf}
\usepackage{amssymb}
\usepackage{amsmath,mathrsfs,verbatim}
\usepackage[usenames, dvipsnames]{color}

\usepackage{times}
\usepackage{latexsym}

\usepackage{floatrow}

\usepackage{longtable}
\usepackage[utf8]{inputenc}
\usepackage{hyperref}
\hypersetup{
    colorlinks,
    citecolor=black,
    filecolor=black,
    linkcolor=black,
    urlcolor=black
}

\def\beq{\begin{equation}}
\def\eeq{\end{equation}}
\def\bey{\begin{eqnarray}}
\def\eey{\end{eqnarray}}

\def\lsim{\mathrel{\raise.3ex\hbox{$<$\kern-.75em\lower1ex\hbox{$\sim$}}}}
\def\gsim{\mathrel{\raise.3ex\hbox{$>$\kern-.75em\lower1ex\hbox{$\sim$}}}}

\begin{document}

\title{Dark Matter Cores and Cusps in Spiral Galaxies and their Explanations}

\author{Manoj Kaplinghat}
\email[]{mkapling@uci.edu}
\affiliation{Department of Physics and Astronomy, University of California, Irvine, California 92697, USA}
\author{Tao Ren}
\email[]{tren001@ucr.edu}
\affiliation{Department of Physics and Astronomy, University of California, Riverside, California 92521, USA}
\author{Hai-Bo Yu}
\email[]{haiboyu@ucr.edu}
\affiliation{Department of Physics and Astronomy, University of California, Riverside, California 92521, USA}
\date{\today}

\begin{abstract}
\vspace*{.0in}
We compare proposed solutions to the core vs cusp issue of spiral galaxies, which has also been framed as a diversity problem, and demonstrate that the cuspiness of dark matter halos is correlated with the stellar surface brightness. We compare the rotation curve fits to the SPARC sample from a self-interacting dark matter (SIDM) model, which self-consistently includes the impact of baryons on the halo profile, and hydrodynamical N-body simulations with cold dark matter (CDM). The SIDM model predicts a strong correlation between the core size and the stellar surface density, and it provides the best global fit to the data. The CDM simulations without strong baryonic feedback effects fail to explain the large dark matter cores seen in low surface brightness galaxies. On the other hand, with strong feedback, CDM simulations do not produce galaxy analogs with high stellar and dark matter densities, and therefore they have trouble in explaining the rotation curves of high surface brightness galaxies. This implies that current feedback implementations need to be modified. We also explicitly show how the concentration-mass and stellar-to-halo mass relations together lead to a radial acceleration relation (RAR) in an averaged sense, and reiterate the point that the RAR does not capture the diversity of galaxy rotation curves in the inner regions. These results make a strong case for SIDM as the explanation for the cores and cusps of field galaxies. 

\end{abstract}

\pacs{95.35.+d}

\maketitle

\section{Introduction}

The cold dark matter (CDM) model is extremely successful in explaining the matter distribution of the Universe on large scales \cite{tegmark2004three} and many important aspects of galaxy formation~\cite{springel2006large,trujillo2011galaxies}. However, it has challenges in matching with observations on galactic scales, see~\cite{Tulin:2017ara,Bullock:2017xww}. Dwarf and low surface brightness galaxies often prefer a cored dark matter density profile rather than a cuspy one predicted in CDM-only simulations~\cite{Flores:1994gz,Moore:1994yx,Persic:1995ru,KuziodeNaray:2007qi,Oh:2010ea,Oh:2015xoa}, which was initially coined as the ``core vs cusp" problem implicitly referring to a conflict between the early predictions of CDM (``cusp") vs observations (``cores"). But, it became clear that observed galaxies are not uniformly cored~\cite{Simon:2004sr} and that the spread in central dark matter densities for similar mass galaxies was hard to explain with existing dark matter models~\cite{McGaugh:2005er,deNaray:2009xj}. More recently, this issue has been reframed as a ``diversity" problem~\cite{Oman:2015xda}, i.e., the inner rotation curves of spiral galaxies exhibit diverse shapes with a spread larger than predicted in the CDM model. In fact, for many dwarf galaxies, there is {\it no} evidence that they inhabit a cored halo. 

In discussing the rotation curves, it is important to keep in mind the uncertainties in inferring them and modeling non-equilibrium and noncircular motion. Simulations~\cite{Pineda2016,read2016understanding,2016ApJ...820..131E,2017ApJ...835..193E,Oman:2017vkl} show that in some cases these could lead to the inference of larger cores than reality and thereby exaggerate the diversity. At present, no conclusive work has demonstrated these effects are at the core of the diversity seen in the rotation curves. At the same time, hydrodynamical CDM simulations have shown that feedback from star formation can modify the inner halo structure and create constant density cores on dwarf scales~\cite{Navarro:1996bv,Governato:2009bg,Teyssier:2012ie,DiCintio:2013qxa,Chan:2015tna,Tollet:2015gqa,Fitts:2016usl,Hopkins:2017ycn,read2016understanding,2018MNRAS.473.4392S}, alleviating some of the tensions, although its significance depends on the feedback model, see~\cite{Sawala:2015cdf,Oman:2015xda,Fattahi:2016nld,Bose:2018oaj}. 

An exciting progress has been the demonstration that the self-interacting dark matter (SIDM) model~\cite{Spergel:1999mh,Kaplinghat:2015aga} provides excellent fits to the diverse rotation curves~\cite{Kamada:2016euw,Creasey:2016jaq,Ren:2018jpt}, while being consistent with CDM predictions on large scales, see~\cite{Tulin:2017ara} for a review. In particular, Ren et al.~\cite{Ren:2018jpt} fitted $135$ galaxies from the SPARC sample~\cite{Lelli:2016zqa} and found that a self-scattering cross section per mass of $\sigma/m=3~{\rm cm^2/g}$ provides an excellent fit for the entire sample. The SIDM model predicts {\it both} cored and cuspy profiles, depending on baryon concentration, and the inferred halo parameters are consistent with the concentration-mass relation from cosmological CDM simulations~\cite{Dutton:2014xda} within the $2\sigma$ range. The resulting stellar mass-to-light ratios are peaked towards $0.5{M_{\odot}/L_\odot}$, consistent with the predictions of the population synthesis models for the $3.6{\mu}m$ band~\cite{McGaugh:2013cca,Schombert:2013hga}. The fits also recover the overall trend of the abundance matching relation as in~\cite{behroozi2013average}. SIDM also provides an explanation for the uniformity of the rotation curves~\cite{Ren:2018jpt}, i.e., the tight correlation between the total and baryonic acceleration scales in galaxies~\cite{McGaugh:2016leg}.

In this paper, we have two aims. The first is to compare the Ren et al.'s fits with those in the literature for CDM simulations. We discuss both the dark matter and baryon distributions inferred from the SIDM fits~\cite{Ren:2018jpt}, and compare them to the predictions of NIHAO~\cite{Tollet:2015gqa,2018MNRAS.473.4392S} and FIRE-2~\cite{Fitts:2016usl,Hopkins:2017ycn} simulations. In these simulations, the feedback is driven by spatially and temporally fluctuating star formation and we label this as ``strong feedback" given its impact on the dark matter distribution. In addition, we compare the $\chi^2/{\rm d.o.f.}$ distributions of the SIDM fits to those of the CDM fits based on the DC14 model~\cite{DiCintio:2014xia}, adapted from Katz et al.~\cite{Katz:2016hyb}. The DC14 model is consistent with halo density profiles from NIHAO simulations~\cite{Tollet:2015gqa,2018MNRAS.473.4392S}. And the general trend for the density profiles recovered in NIHAO simulations is similar to that in FIRE-2 simulations, hence our conclusions would apply to both. We also consider models where the feedback is driven by spatially smoother star formation that does not lead to dark matter cores~\cite{Sawala:2015cdf,Oman:2015xda,Fattahi:2016nld,Bose:2018oaj}. In these CDM simulations with ``weak feedback", the Navarro-Frenk-White (NFW) profile~\cite{Navarro:1995iw,Navarro:1996gj} provides a reasonable approximation for halo density profiles on dwarf scales~\cite{Oman:2015xda}. Keeping this in mind, we also include fits using the NFW profile from~\cite{Katz:2016hyb}. 

Our second aim is to compare the SIDM and CDM fits with the fits using the radial acceleration relation (RAR)~\cite{Li:2018rnd}. This empirical relation is based on the average over the SPARC sample in the plane of the total radial vs baryon accelerations~\cite{McGaugh:2016leg}. We further propose an {\it one-parameter} SIDM model that has only one variable, i.e., the mass-to-light ratio, in fitting the data, as in the case of the RAR. And we use it to highlight the importance of diverse dark matter distributions in explaining stellar kinematics of spiral galaxies.  

The plan for this paper is as follows. In Sec. II, we first provide a visual representation of the key issue that we will identify for current CDM simulations with strong baryonic feedback. We use a small subset of galaxies from the SPARC sample to show that the cuspiness of the dark matter halo is related to the compactness of the stellar distribution and illustrate the challenge for the CDM simulations in reproducing this pattern. 

In Sec. III, we discuss the inferred logarithmic slope of the dark matter density profile at $1.5\%$ of the virial radius from the SIDM fits and compare it with the predictions in NIHAO and FIRE-2 simulations. Our analysis indicates that the CDM simulations with strong feedback do not create galaxies that are as cuspy as those predicted by SIDM. We show that the inferred inner slope actually correlates with the stellar surface density in the sense that large cores are recovered within SIDM for low surface brightness galaxies. In Sec. IV, we present the cumulative distributions of $\chi^2/{\rm d.o.f.}$ values inferred from SIDM and CDM fits and show the SIDM model is by far the best description of the data. 

In Sec. V, we perform a new SIDM fit to the SPARC sample, but without including scatters in the priors. In this limit, the SIDM model and the RAR have the same variable in the fits, i.e., the stellar mass-to-light ratio. We show even in this extreme case the SIDM fits are better than the RAR fits and demonstrate that the one-parameter SIDM fits do not lead to one-to-one correlation between the total radial and baryonic accelerations. We conclude in Sec. VI.

\section{Understanding the core vs cusp problem and its correlation with the stellar density}
\label{sec:feedback}

\begin{figure}[t!]
\centering
 \includegraphics[scale=0.45]{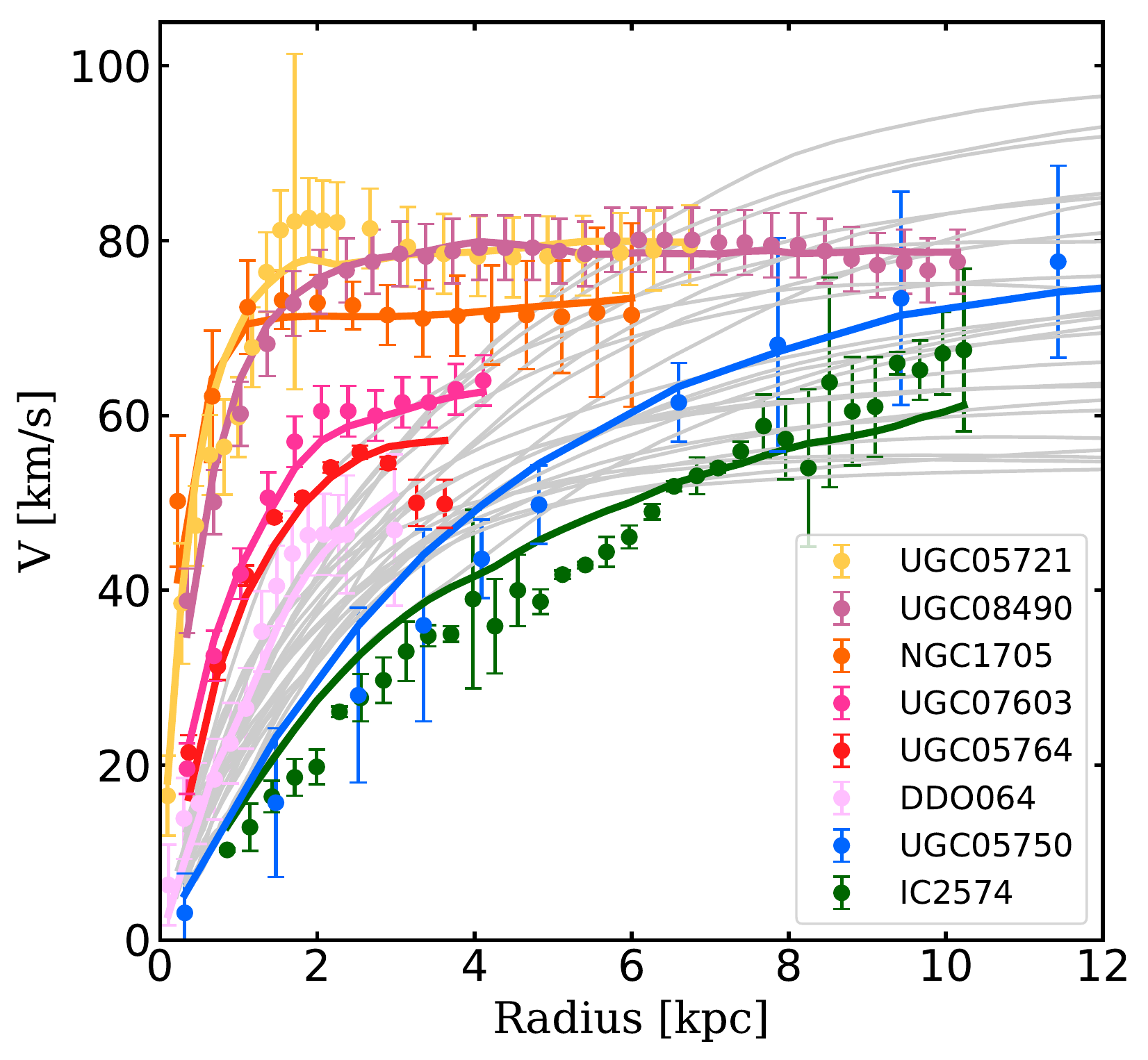}\;\;\;\;
 \includegraphics[scale=0.45]{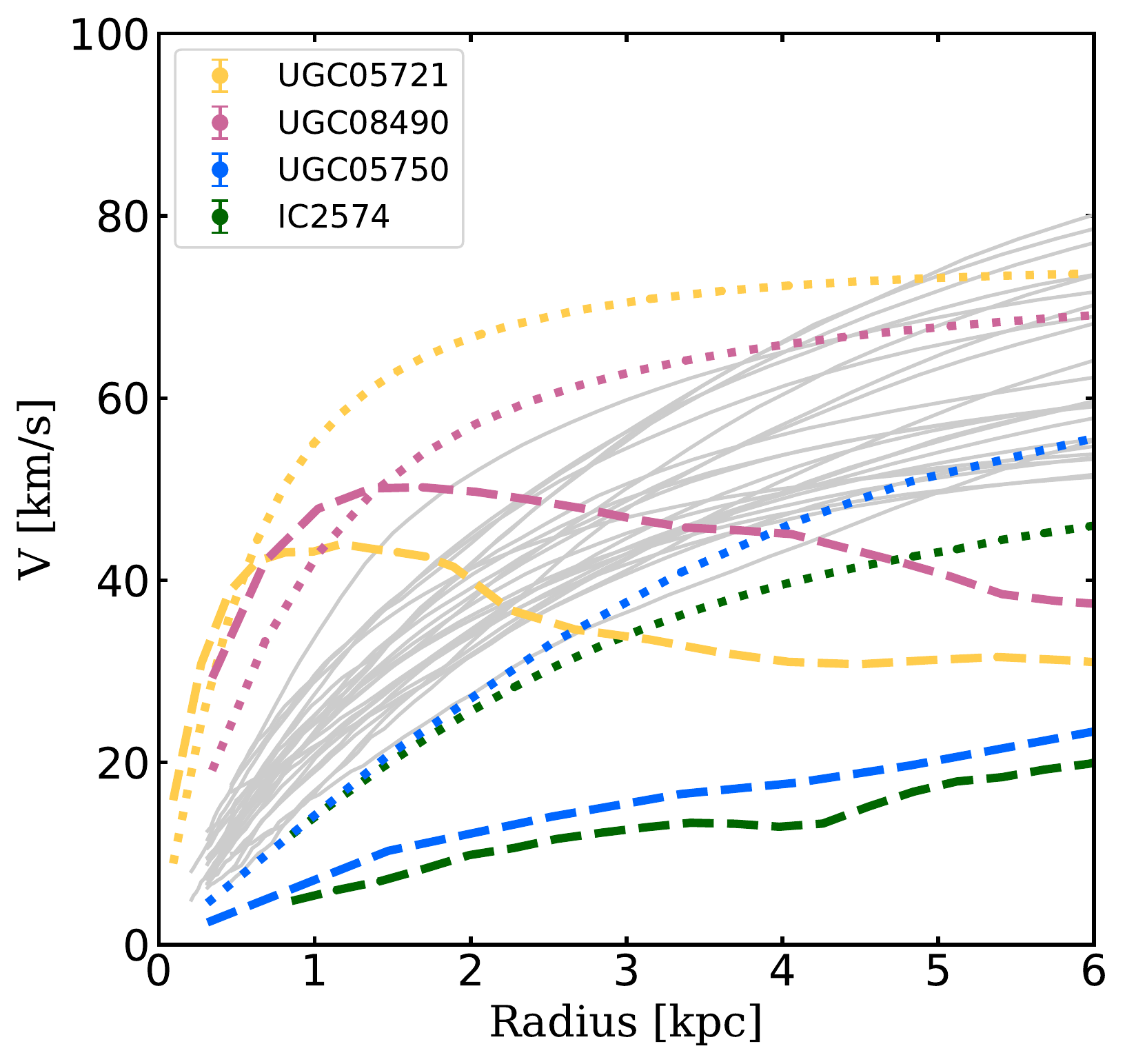} 
\caption{{\it Left:} SIDM fits (solid-colored) to the rotation curves of eight galaxies in the SPARC sample (colored dots with error bars), spanning the full range of the diversity, compared with simulated analogs in a similar velocity range from NIHAO hydrodynamical CDM simulations with strong feedback (gray)~\cite{2018MNRAS.473.4392S}. {\it Right:} Decomposed halo (dotted) and baryonic (dashed) contributions to the total rotation curves for the four outliers, together with the simulated total rotation curves as shown in the left panel (gray).}
\label{fig:outliers}
\end{figure}

Compared to CDM, the SIDM model has one additional parameter, the dark matter self-scattering cross section per mass ($\sigma/m$). The overall fits of the galaxy sample are not sensitive to a specific value of the cross section as long as $\sigma/m\sim{\cal O}(1)~{\rm cm^2/g}$~\cite{Ren:2018jpt}, so it is hard to use the rotation curve data to pin down $\sigma/m$. For this range of the cross section, the SIDM model predicts {\it both} cored and cuspy profiles, depending on baryon concentration. For galaxies with low baryon concentration, dark matter self-interactions thermalize the inner halo and produce a density core over the age of galaxies~\cite{Dave:2000ar,Vogelsberger:2012ku,Zavala:2012us,Rocha:2012jg,Peter:2012jh,Elbert:2014bma}. While for those with higher baryon concentration, the SIDM thermalization in the presence of the baryonic potential actually leads to a dense inner halo with a smaller core~\cite{Kaplinghat:2013xca}. Thus, the ``cuspiness" of the SIDM halo is correlated with the baryon concentration, and this prediction is robust to detailed formation histories of galaxies as long as the cross section is large~\cite{Elbert:2016dbb,Robertson:2017mgj,Sameie:2018chj}.

In fitting the rotation curves of the SPARC sample, Ren et al. assumed the inner halo is thermalized due to strong self-interactions and it follows a cored isothermal density profile characterized by the central dark matter density ($\rho_0$) and the one dimensional dark matter velocity dispersion ($\sigma_{\text{v0}}$)~\cite{Kaplinghat:2015aga, Kamada:2016euw}. The inner isothermal profile is matched to a NFW profile at $r_1$, where SIDM particles scatter once over the age of galaxies. This simple procedure to match the solutions in the inner and outer regions works well to describe the SIDM halos in a variety of simulations~\cite{Kaplinghat:2015aga,Ren:2018jpt}. 

There is one-to-one correspondence between the inner halo parameters ($\rho_0$, $\sigma_{\rm v0}$) and the NFW halo ones ($V_{\rm max}$, $r_{\rm max}$), i.e., the maximal circular velocity and its associated radius. Thus, the number of free parameters for a halo is two, the same as CDM fits. Ren et al. considered two independent fitting approaches. The controlled sampling (CS) uses an axisymmetric exponential thin disk to model the stellar disk~\cite{Kamada:2016euw} and constrained the $V_{\rm max}\textup{--}r_{\rm max}$ relation within the $2\sigma$ range predicted in cosmological simulations~\cite{Dutton:2014xda}. The Markov Chain Monte Carlo sampling (MS) assumes spherical symmetry for the baryon distribution and explored the full likelihood of the parameter space. The MS approach imposes a constraint on the $V_{\rm max}\textup{--}r_{\rm max}$ relation, such that the fits have to be within the $3\sigma$ range. The resulting median concentrations for almost all the galaxies are within the $2\sigma$ range, in agreement with the CS approach. Both approaches provide excellent fits to the data and the results agree each other well. In addition, the stellar and halo masses are strongly correlated, in agreement with the abundance matching relation~\cite{Behroozi:2012iw}. This not imposed in either the CS or the MS approach, and hence an important test of the consistency of the SIDM model.

Santos-Santos et al. showed hydrodynamical NIHAO CDM simulations with strong baryon feedback can successfully reproduce the overall trend of diverse galaxy rotation curves in the SPARC sample~\cite{2018MNRAS.473.4392S}. There are a few outliers that the simulations do not match well. In particular, they highlighted $11$ such galaxies whose last-measured circular velocities are in the range of $47\textup{--}90~{\rm km/s}$, while their circular velocities at $2~{\rm kpc}$, $V (2{\rm kpc})$, are $\pm3\sigma$ away from the simulation average. We consider eight of them, which are in the SPARC sample and span the full range of the diversity, for a more detailed comparison. They are UGC 05721, UGC 08490, NGC 1705, UGC 07603, UGC 05750, DDO 064, UGC 05750 and IC 2574. We use these galaxies to exemplify key points because they have been used similarly in the literature. Our final conclusions do not hinge on these eight galaxies; it is based on fitting to more than a hundred rotation curves. 

Fig.~\ref{fig:outliers} (left) shows observed rotation curves for the eight outliers (colored dots with error bars), together with SIDM fits (solid-colored) and the simulated rotation curves in the mass range (gray). 
It is clear that these outliers have diverse profiles with almost a factor of $4$ spread in $V(2~{\rm kpc})$ -- IC 2574 and UGC 05750 have the lowest $V(2~{\rm kpc})$, $20~{\rm km/s}$, while UGC 08490 and UGC 05721 have the highest, $80~{\rm km/s}$. Despite the spread, the SIDM model can provide good fits. For galaxies like IC 2574 and UGC 05750, the SIDM halos have low concentration~\cite{Kamada:2016euw,Ren:2018jpt}. In addition, as indicated in Fig.~\ref{fig:outliers} (right), dark matter completely dominates over the baryons in these galaxies and self-interactions lead to large density cores in accord with observations. For UGC 05721 and UGC 08490, the baryon concentrations are much higher, and SIDM thermalization in the presence of the deeper baryonic potential leads to denser and smaller dark matter cores, as shown in Fig.~\ref{fig:outliers} (right). 

In NIHAO simulations, baryonic feedback is strong enough to cause core expansion in the inner regions of the halo. As shown in Fig.~\ref{fig:outliers} (left), the inner rotation curves of the simulated galaxies have systematically slower rise than expected from CDM simulations with weak feedback~\cite{Oman:2015xda}. Thus, while NIHAO simulations can explain IC 2574 and UGC 05750, they do not produce analogs of galaxies with sharply rising rotation curves; see also~\cite{Ghari:2018jkc}. The simulated curves are close to being consistent with DDO 064 and UGC 05764, but too shallow to explain UGC 05721, UGC 08490 and UGC 7603, which are more consistent with the NFW profile. The predicted spread in $V(2~{\rm kpc})$ is a factor of $2$ for the simulated galaxies.
Note the $\pm2 \sigma$ scatter in the halo concentration-mass relation~\cite{Dutton:2014xda} can lead to a factor of $2$ spread in $V(2~{\rm kpc})$ for a halo with $V_{\rm max}=70~{\rm km/s}$~\cite{Kamada:2016euw}. So, it seems that including baryonic feedback shifts the median but does not increase the scatter.

From the eight galaxies we have highlighted, there is a clear pattern that the ``cuspiness" of the halo is correlated with the baryon distribution, which was recognized more than a decade ago~\cite{McGaugh:2005er}. Galaxies that favor large density cores in dark matter also have low densities of stars, while those with higher stellar densities are compatible with cuspy halos. Both types of galaxies are represented in the field population. A viable solution to the diversity problem must lead to cores and cusps that are correlated correctly with the observed stellar distribution, which we discuss further later. 

Hydrodynamical CDM simulations with the strong feedback can produce cored halo profiles, but they do not seem to create galaxies with stellar sizes on the compact end of the observed distribution. This point has been noted for the FIRE-2 dwarf galaxy simulations~\cite{2019MNRAS.487.1380G}. A possible explanation is that if feedback is significant enough to push CDM particles to larger orbits and reduce the central densities, it would do the same for stars because stars and dark matter are both collisionless in CDM simulations. On the other hand, for CDM simulations with weak feedback effects, the halo remains cuspy and it is hard to produce extended stellar distributions seen in low surface brightness galaxies.  It remains to be seen whether alternate feedback prescriptions that can populate the full range of observed stellar and dark matter densities within the CDM model. 

It is important to note that some form of feedback that pushes gas around is necessary to obtain the stellar-to-halo mass relation and thereby explain the inefficiency of galaxy formation for stellar masses below about $10^{10}~\rm M_\odot$. This is true regardless of the nature of dark matter. 
The critical point we are highlighting here is that the stellar disk sizes provide an independent probe of the validity of these tuned feedback models, and the impact on dark matter densities and disk sizes are correlated. We discuss this in more detail next.

\section{The diversity of the inner dark matter density profiles}

\begin{figure}[t]
\includegraphics[scale=0.45]{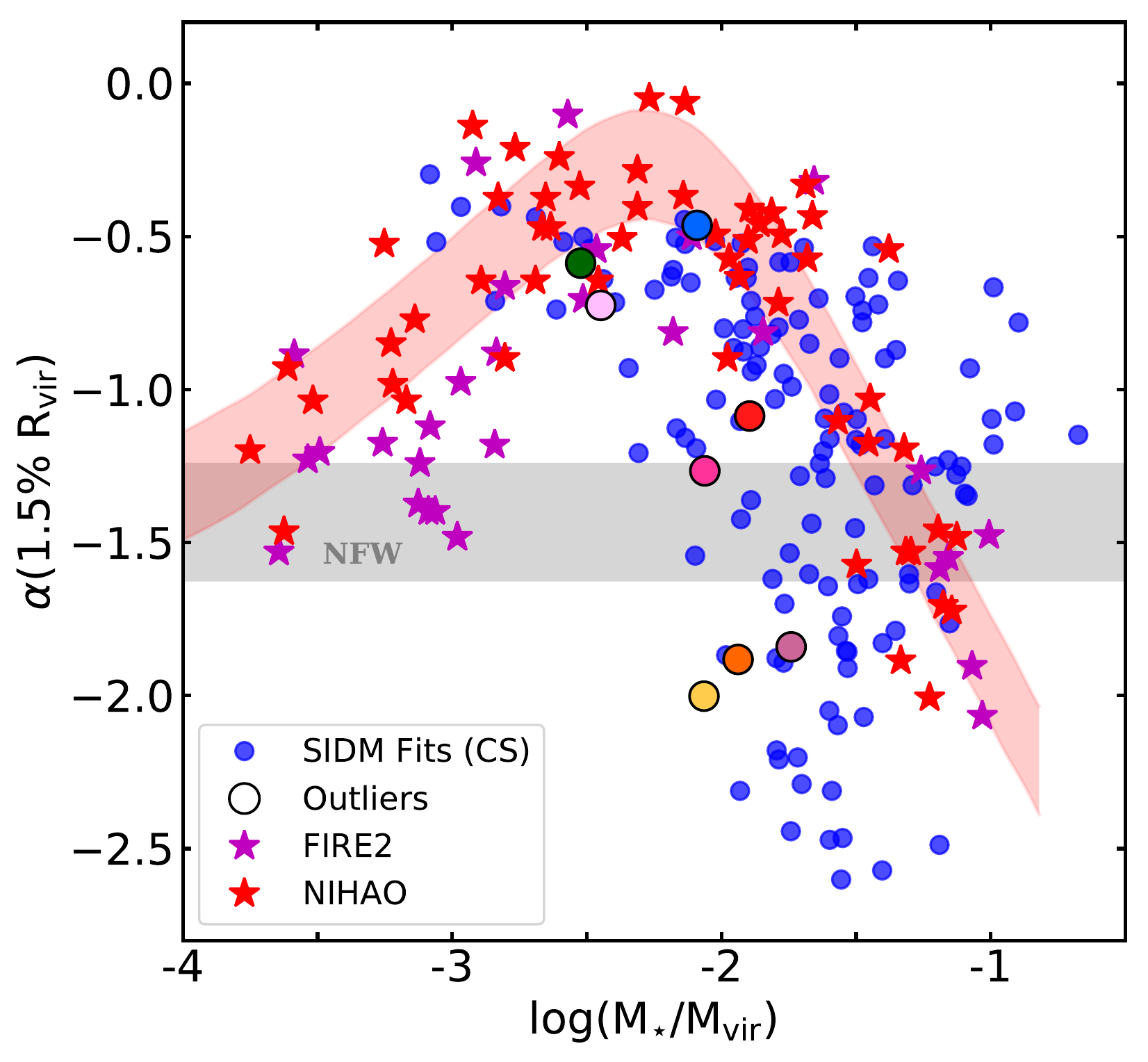} \;\;\;\;
\includegraphics[scale=0.45]{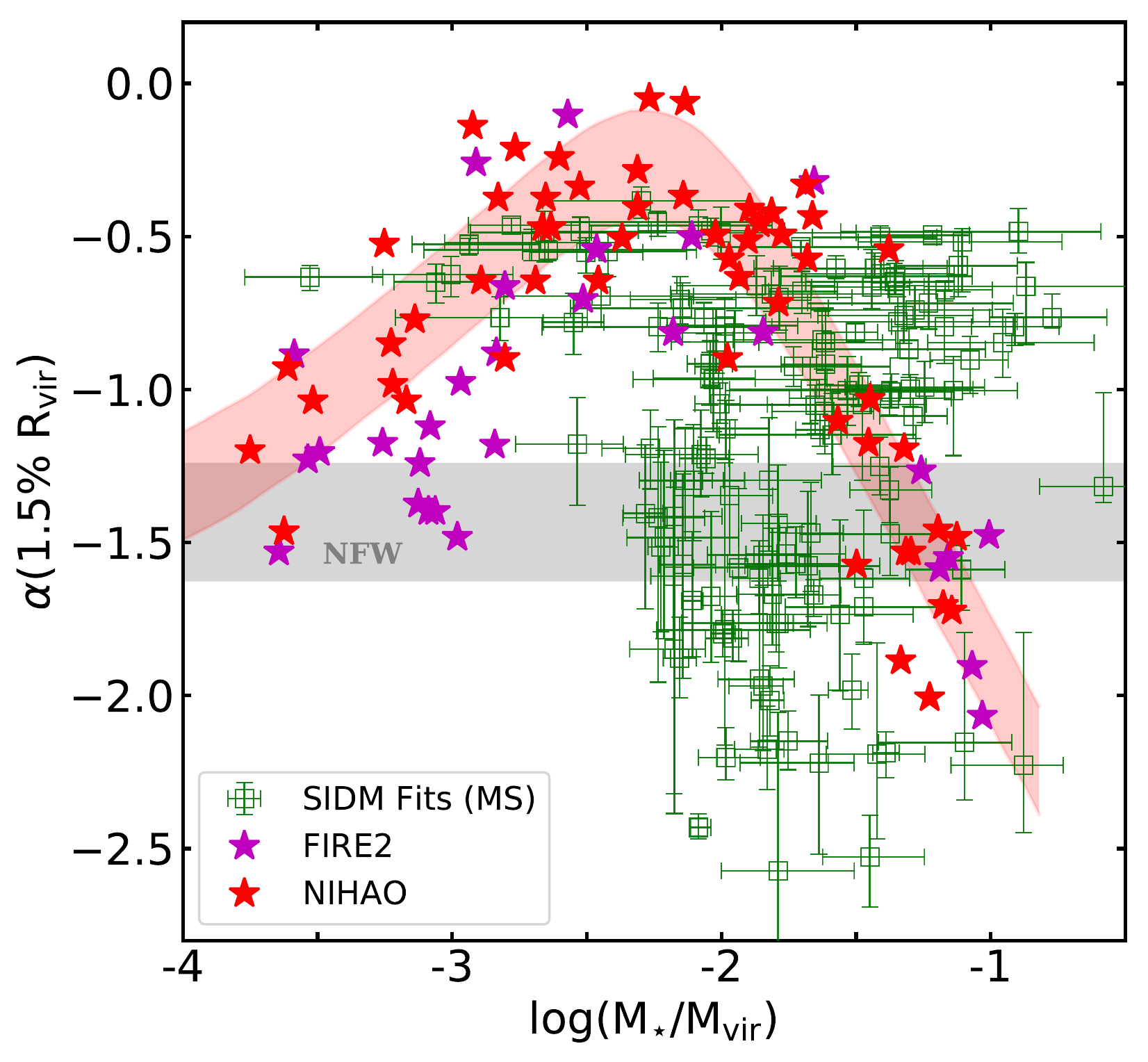}
\caption{{\it Left:} Logarithmic slope of the dark matter density profile at $1.5\% r_{\rm vir}$ vs the ratio of stellar-to-halo masses inferred from the SIDM fits with controlled sampling in Ren et al.~\cite{Ren:2018jpt}. Large filled circles denote the outliers shown in Fig. 1 with the same color scheme. For comparison, we also show the expected range from NIHAO~\cite{Tollet:2015gqa} (red band to guide the eye) and FIRE-2~\cite{fitts2017fire,hopkins2018fire} hydrodynamical CDM simulations, as well as CDM-only simulations (gray band); adapted from~\cite{Bullock:2017xww}. {\it Right:} The same as the left panel, but with the SIDM fits using MCMC sampling.}
\label{fig:slope1}
\end{figure}

To be more quantitative, we use the logarithmic slope of the density profile at $1.5\%$ times the virial radius $r_{\rm vir}$, $\alpha$, to characterize the cuspiness of the halo in the SIDM fits, study its correlations with the ratio of stellar-to-halo masses ($M_*/M_{\rm vir}$) and the central stellar mass density ($\Sigma_0=M_*/2\pi R^2_{\rm d}$), and compare them with CDM simulations. 

Fig.~\ref{fig:slope1} shows the logarithmic slope vs $\log (M_*/M_{\rm vir})$ interfered from the SIDM fits with controlled (left) and MCMC (right) samplings~\cite{Ren:2018jpt}, together with the results from NIHAO~\cite{Tollet:2015gqa} and FIRE-2~\cite{fitts2017fire,hopkins2018fire} simulations, as well as CDM-only predictions, adapted from~\cite{Bullock:2017xww}. We use large filled circles to denote outlier galaxies shown in Fig.~\ref{fig:outliers}. We first note that $\alpha$ from the SIDM fits spans a large range from $-0.5$ to $-2.5$, indicating that the SIDM model predicts both cored {\it and} cuspy inner halos. In fact, about $50\%$ of the galaxies in our SIDM fits have $\alpha\lesssim-1$. For $\log(M_*/M_{\rm vir})\sim-1.5$, $\alpha$ has the largest spread. This reflects diverse baryon distributions in the galaxies even their stellar-to-halo mass ratios are fixed, as the SIDM halo profile is very responsive to the baryonic potential, $\rho_{\rm iso}\propto\exp(-\Phi_{\rm b}/\sigma^2_{\rm v0})$~\cite{Kaplinghat:2013xca,Elbert:2016dbb,Sameie:2018chj,Robertson:2017mgj}. We will come back to this point later.

On the other hand, both NIHAO and FIRE-2 CDM simulations predict a similar trend. For $\log (M_*/M_{\rm vir})\lesssim-3.5$, the baryon content is too small to change the inner halo structure and the profile remains cuspy. The halo becomes more core-like as  $M_*/M_{\rm vir}$ increases due to the strong feedback, the maximal core expansion occurs when $\log (M_*/M_{\rm vir})\sim-2.5$. The contraction effect starts to dominate and the halo becomes cuspy again when the mass ratio increases further. For many dwarf galaxies with density cores, the slopes inferred from the SIDM fits are consistent with the predictions in the CDM simulations such as the outliers DDO 064, UGC 05750 and IC 2574. However, overall the SIDM fits exhibit a much larger spread in $\alpha$ for a given ratio of $M_*/M_{\rm vir}$. For galaxies with $\log(M_*/M_{\rm vir})\sim-1.5$, $\alpha$ varies from $-2.5$ to $-0.5$ in the both SIDM fits, while both CDM simulations predict $\alpha$ in a much smaller range from $-1.0$ to $-0.7$. Among the eight outliers shown in Fig.~\ref{fig:outliers}, four of them have $\alpha$ below $-1.5$ in the SIDM fits (UGC05721, UGC 08490, NGC1705). According to the CDM simulations, their $\alpha$ values would be larger than $-1.0$ given their $M_{\star}/M_{\rm vir}$ ratios. Thus, to fit their rotation curves, one expect that the simulations would produce higher baryon concentration compared to the one inferred from the SIDM fits to these galaxies. However, this is not the case as shown in Fig.~\ref{fig:outliers} (right). Another option would be to increase the scatter of CDM in the horizontal direction. For example, we could relax the abundance matching relation imposed in the simulations and shift $M_{\rm vir}$ for given $M_\star$. From Fig.~\ref{fig:slope1}, we see that the required variation is about $\pm0.5$ dex and it is interesting to see whether the CDM simulations can reproduce the scatter. Note the SIDM fits~\cite{Ren:2018jpt} recover the trend of the abundance matching interference as in~\cite{Behroozi:2012iw}.

\begin{figure}[t]
\centering
\includegraphics[scale=0.45]{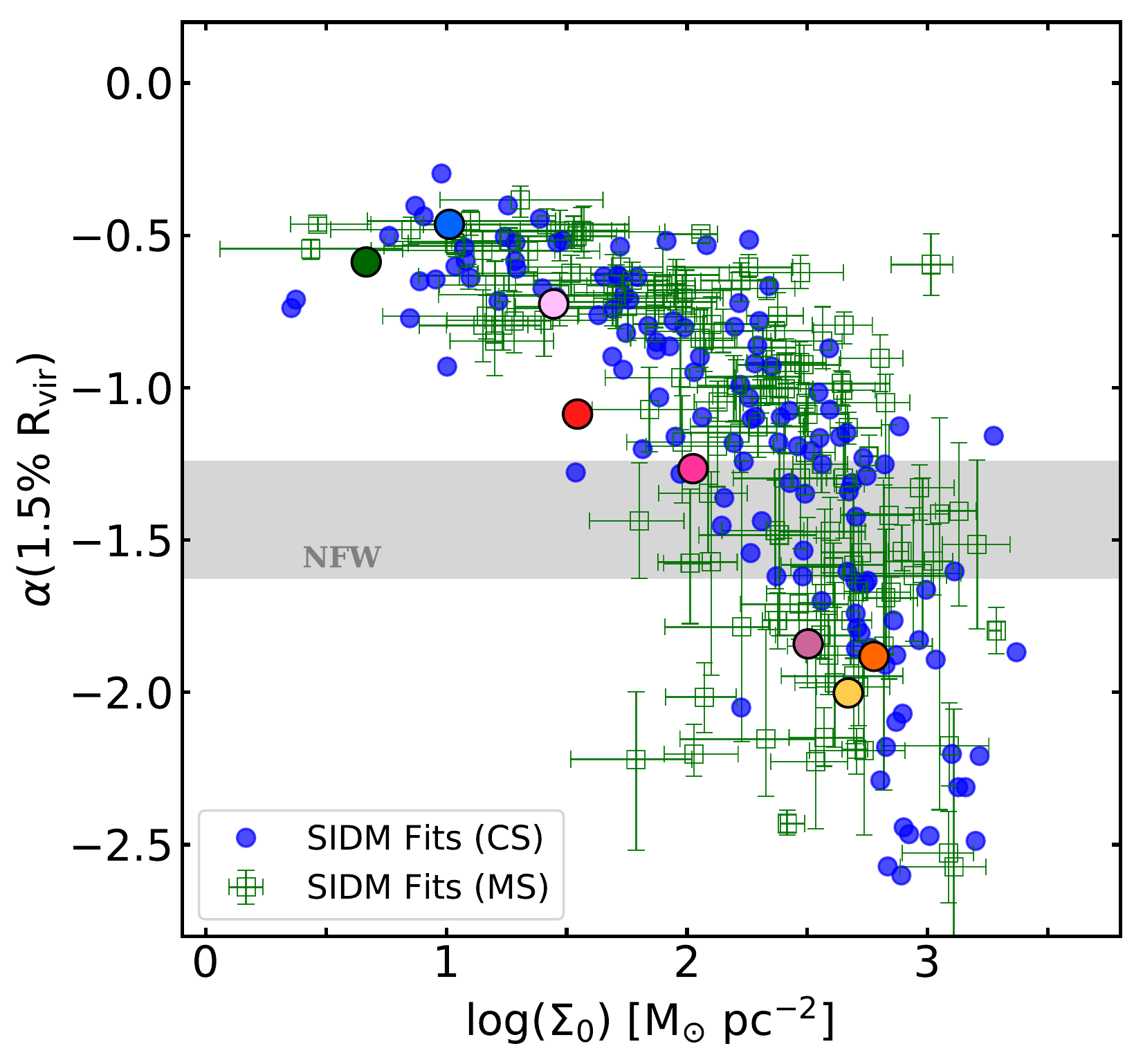} \;\;\;\;
\includegraphics[scale=0.45]{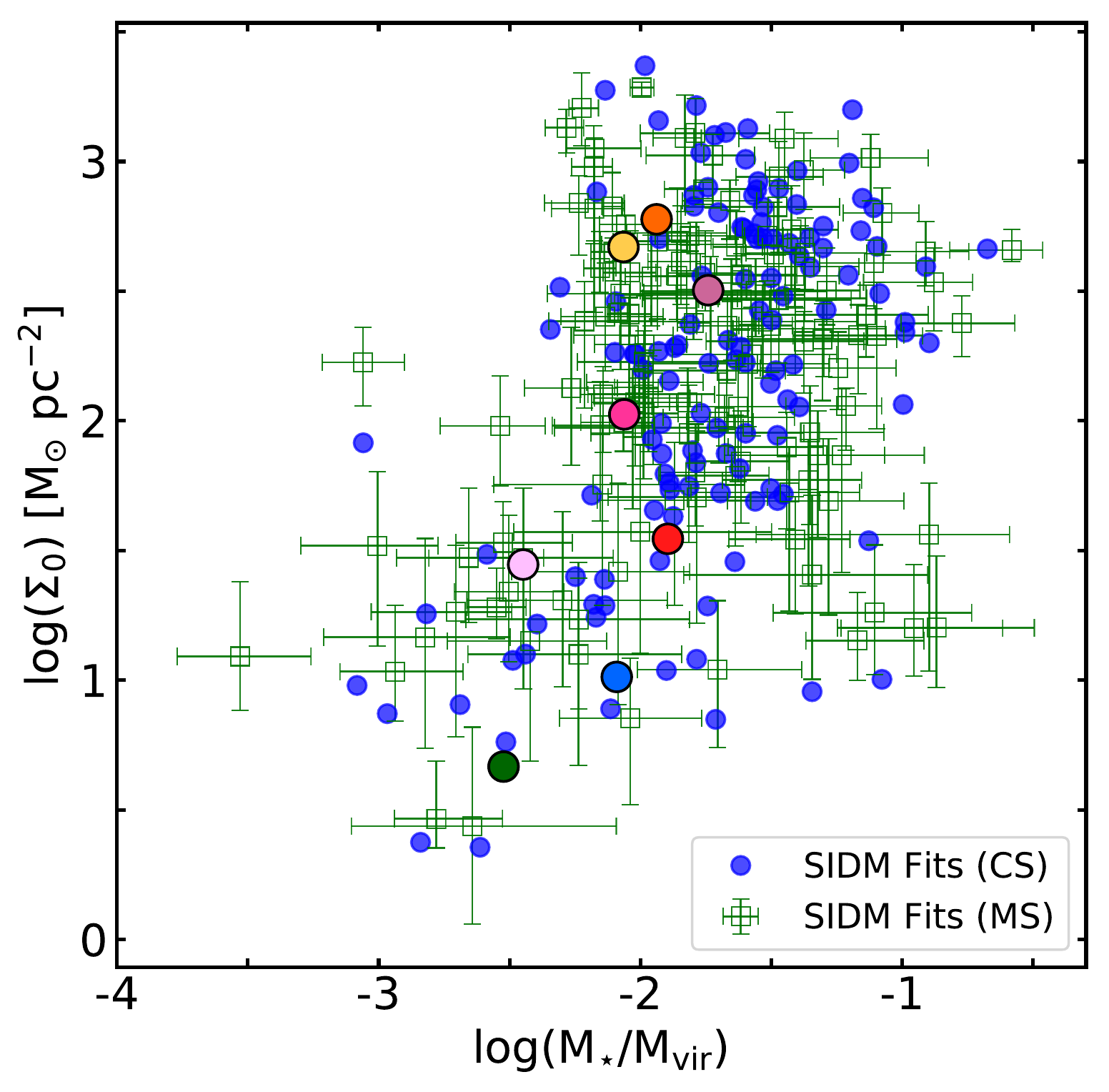} 

\caption{{\it Left:} Logarithmic slope of the halo density profile vs the central stellar surface density, inferred from the SIDM fits with controlled (blue) and MCMC (green) samplings in Ren et al.~\cite{Ren:2018jpt}. {\it Right:} The corresponding central stellar surface density vs the ratio of stellar-to-halo masses. In both panels, large filled circles denote the outliers shown in Fig.~\ref{fig:outliers}.}
\label{fig:slope2}
\end{figure}

Fig.~\ref{fig:slope2} (left), we plot the logarithmic slope vs the stellar mass surface density. There is a clear pattern that $\alpha$ is strong correlated with $\Sigma_0$. For $\Sigma_0\lesssim100~{\rm M_\odot~kpc^{-2}}$, the baryonic influence on the halo is small and SIDM thermalization produces large density cores that are required to match with observations. When $\Sigma_0$ becomes larger, the core size shrinks and the density increases accordingly. This is exactly what is observed, i.e., no large constant density cores in high surface brightness galaxies. Note the scatter in the $\alpha\textup{--}\Sigma_0$ relation shown in Fig.~\ref{fig:slope2} (left) is expected because the baryonic potential is not completely determined by $\Sigma_0$ and there is a spread in $\sigma_{\rm v0}$, on which  $\rho_{\rm iso}$ depends, at fixed $\Sigma_0$.

One point worth emphasizing is that $M_\star/M_{\rm vir}$ does not determine the properties of spiral galaxies, in particular the central surface density as shown in Fig.~\ref{fig:slope2} (right). Overall, $\Sigma_{0}$ increases with $M_*/M_{\rm vir}$, but the spread in $\Sigma_0$ can be a factor of $10\textup{--}100$ for given the mass ratio. This implies that the scale radius of the disk can differ by a factor of $10$ for a given halo mass, up to the scatter in the abundance matching relation. The key challenge for all structure formation simulations is to reproduce the large spread in the stellar density for a given $M_\star/M_{\rm vir}$. For example, recent FIRE-2 simulations~\cite{Fitts:2018ycl} show that there is a tight correlation between the stellar half-mass radius and the stellar mass for both SIDM and CDM runs with halo masses of $10^{10}~{\rm M_\odot}$ and stellar masses in the range of $10^4\textup{--}10^7~{\rm M_\odot}$ at $z=0$.

In the SIDM model, we expect that requirements for the success are different from those in the CDM model. For example, the process of core creation due to dark matter self-interactions can directly lead to adiabatic expansion of the stellar disk~\cite{Vogelsberger:2014pda}. In this case, dwarf galaxies with extended disks may form even if the feedback is weak. At the same time, it has to be strong enough such that stellar bulges are not overproduced, a key motivation for some of the early feedback implementations in CDM simulations, see, e.g.,~\cite{Governato:2012fa}. Thus, the coupling between feedback and the change in the potential well due to SIDM core creation may provide a point of departure from CDM predictions.

More observational input directly related to tracers of star formation and gas may help refine feedback models~\cite{Semenov:2018lku}. Indeed, recent zoom-in simulations~\cite{2019MNRAS.487.1717F}, implemented with strong-feedback models, show that the predicted correlation between tracers of star formation and molecular gas on sub-100-pc scales is too strong to be consistent with observations~\cite{Kruijssen:2014vsa}, while maintaining good agreement with larger-scale observables. A possible resolution is that pre-supernova feedback is stronger than the one implemented~\cite{2019MNRAS.487.1717F}, which could impact the way gas outflows change the gravitational potential well. Given these considerations, it is important to reassess feedback recipes in the context of both CDM and SIDM models, and confront them with the challenge of explaining the wide range of stellar surface brightnesses.

It is also interesting to see whether other well-studied dark matter models, such as warm dark matter (WMD)~\cite{Dodelson:1993je,Shi:1998km,Dolgov:2000ew,Bode:2000gq,Abazajian:2001nj,Asaka:2005an} and fuzzy dark matter (FDM)~\cite{Hu:2000ke,Marsh:2013ywa,Schive:2014dra,Hui:2016ltb}, can reproduce diverse dark matter density profiles along with observed baryon distributions. WDM halos are less dense than their CDM counterparts due to suppressed power spectra, but their density profiles remain cuspy, see, e.g.,~\cite{Lovell_2017,Bozek:2018ekc,Fitts:2018ycl}. FIRE-2 WDM simulations show that the baryonic feedback can produce density cores in WDM halos in a way similar to the case of CDM~\cite{Fitts:2018ycl}. Thus, we expect that WDM has a trend in the $\alpha\textup{--}\log(M_*/M_{\rm vir})$ plane similar to the CDM one shown in Fig.~\ref{fig:slope1} (magenta), although the former predicts more diverse star formation histories~\cite{Bozek:2018ekc}. 

FDM predicts a solitonic core in the inner halo due to the interplay between gravity and quantum pressure~\cite{Schive:2014dra,Schive:2014hza,Marsh:2015wka,Mocz:2017wlg}, and the core size is comparable to the de Broglie wavelength of the FDM particle. For the FDM model with the particle mass of $\sim10^{-22}~{\rm eV}$, the predicted cores can be consistent with observations in dwarf spheroidal galaxies in the Milky Way~\cite{Chen:2016unw} and field dwarfs with halo mass  $\sim10^{10}~{\rm M_\odot}$~\cite{Robles:2018fur}. The FDM halo core size decreases with increasing halo mass, which is at odds with the fact that the largest cores are not observed in the dwarf galaxies but more massive low surface brightness galaxies. Given the current understanding, the FDM model does not explain the diverse rotation curves of more massive galaxies such as those in the SPARC sample~\cite{Robles:2018fur,Bar:2018acw}.  
Unless baryonic feedback can alter the solitonic structure in accord with observations, FDM cannot explain stellar kinematics of galaxies across a range of mass scales. 

\section{Statistical comparisons}

\begin{figure}[t]
\centering

\includegraphics[scale=0.5]{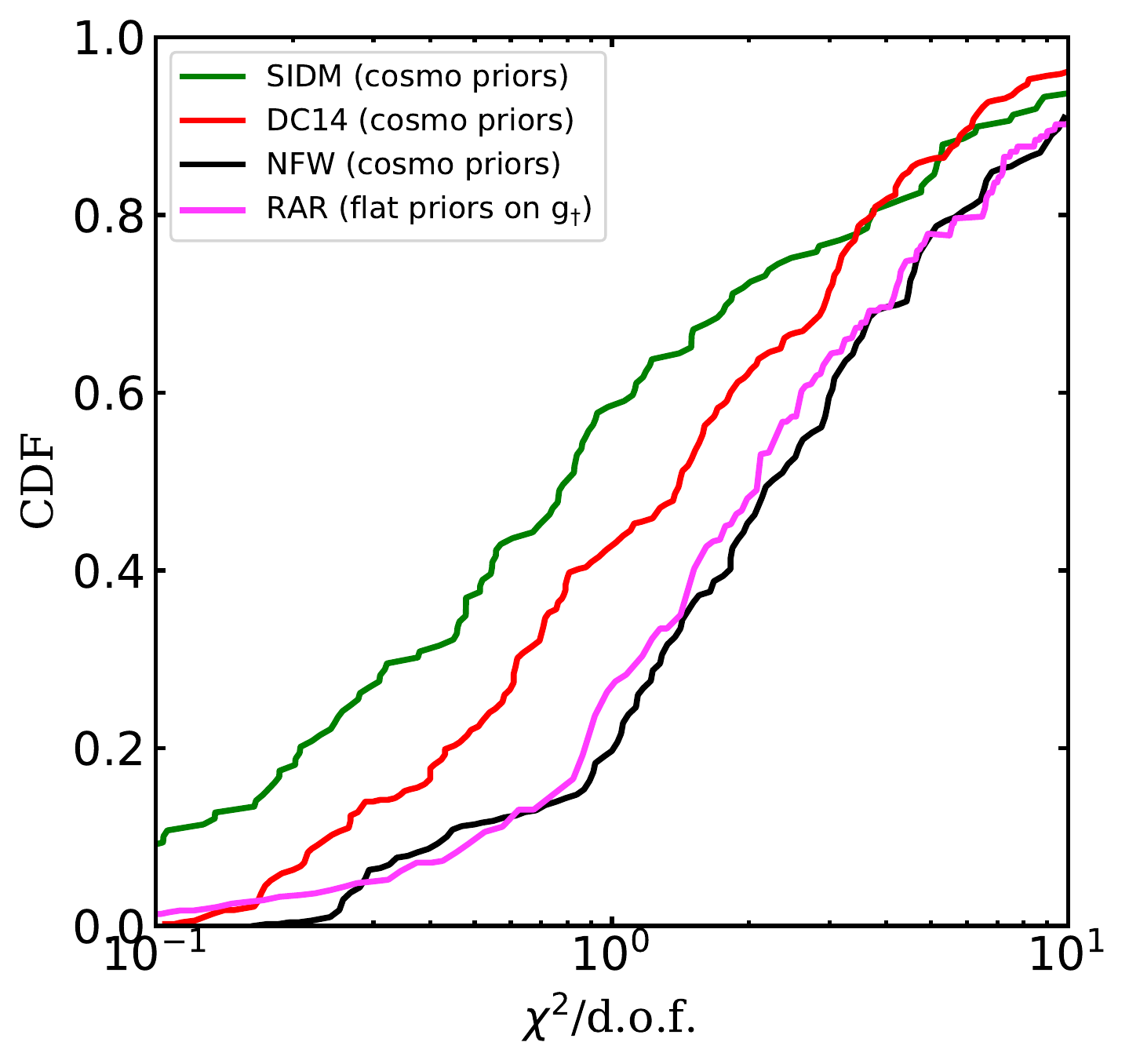} 
\caption{The cumulative distribution functions of $\chi^2/{\rm d.o.f.}$ values for the maximum posterior rotation curve fits of the SIDM model, reprocessed based on the MCMC fits in Ren et al.~\cite{Ren:2018jpt}, the CDM model with strong baryonic feedback (DC14) and the NFW model, adapted from Katz et al.~\cite{Katz:2016hyb}, and the RAR from Li et al.~\cite{Li:2018rnd}. }
\label{fig:cdf}
\end{figure}

After having examined the dark matter and baryon distributions for individual galaxies, we turn to statistical comparisons between SIDM and CDM fits, as well as the ones using the RAR. Katz et al.~\cite{Katz:2016hyb} considered $147$ SPARC galaxies and fitted their rotation curves using the halo profiles based on the NFW profile and CDM simulations with strong feedback (DC14)~\cite{DiCintio:2014xia,DiCintio:2013qxa}. The latter predict CDM density profiles similar to those found in NIHAO simulations~\cite{Tollet:2015gqa}. The Katz et al. sample is based on the following cuts in the SPARC sample: quality factor $Q<3$, inclination angles of $30$ degrees or higher and at least $5$ data points along the rotation curve. When this cut is made, we get $149$ galaxies, two more than Katz et al.'s sample as they used an earlier version of the SPARC sample~\cite{misc}. Among these $149$ galaxies, $123$ of them have been included in Ren et al.'s original fits. For these galaxies, we have post-processed the MCMC chains to only keep the points with the stellar mass-to-light ratio, $\Upsilon_{\star}$, in the range of $0.3 < \Upsilon_{\star}/(M_{\odot}/L_{\odot})< 0.8$ for both the disk and the bulge (when presents), to be consistent with the assumption in~\cite{Katz:2016hyb}, and the minimum $\chi^2/{\rm d.o.f.}$ is computed with this constraint. For the additional $23$ galaxies, we have performed new SIDM fits as in Ren et al., but changed the constraint on $\Upsilon_\star$ to $0.3 < \Upsilon_{\star}/(M_{\odot}/L_{\odot})< 0.8$. 

Fig.~\ref{fig:cdf} shows the cumulative distribution functions of $\chi^2/{\rm d.o.f. }$ values for the maximum posterior rotation curve fits of the SIDM model in Ren et al.~\cite{Ren:2018jpt}, as well as the CDM model with strong baryonic feedback and the NFW model, adapted from Katz et al.~\cite{Katz:2016hyb}. Note we show the results for two cases from~\cite{Katz:2016hyb}: the fits with DC 14 and NFW models after imposing lognormal priors on the halo concentration-mass relation as in~\cite{Dutton:2014xda} and the abundance matching relation in~\cite{Moster:2012fv}. For reference, we have also included the results from Li et al.~\cite{Li:2018rnd} on the fits using the RAR, and we will discuss them in the next section, Sec. V.

The DC14 model is better than the NFW model in fitting the data because of core formation driven by the baryonic feedback, as illustrated in Fig.~\ref{fig:slope1} (left). We also see that SIDM fits are superior to the DC14 fits. Both fits have two degrees of freedom for the halo, essentially $V_{\rm max}$ and $r_{\rm max}$, and they also impose constraints on them using the cosmological concentration-mass relation~\cite{Dutton:2014xda}. In the case of the DC14 fits of Katz et. al., the prior is lognormal with a spread of $0.1$ dex. We note, however, that a substantial fraction of the inferred concentrations tend to scatter well above the $2\sigma$ band; see their Fig. 5. While for the SIDM fits, Ren et. al. used a top-hat constrain on $\log_{10}c_{200}$ with a width of $\pm 0.33$ dex. The inferred concentrations are predominantly within about $0.2$ dex, see Fig. 3 in~\cite{Ren:2018jpt}. 

In addition, Katz et al. also imposed an abundance-matching prior relating the stellar to halo masses in their fits. This prior was not imposed in Ren et al.'s SIDM fits, but the recovered stellar and halo masses show a strong correlation for the full sample, which is in excellent agreement with the abundance matching relation in~\cite{Behroozi:2012iw} in the high mass regions where it was measured. Thus, the abundance matching expectations fall out of the SIDM fits.

\section{Fits with the radial acceleration relation}

Lastly, we consider the fits to SPARC galaxies with the RAR. It assumes that the galactic rotation curve is fully determined by the stellar and gas distribution and an acceleration scale that is common to all galaxies~\cite{McGaugh:2016leg}. In Fig.~\ref{fig:cdf}, we show the cumulative distribution function of $\chi^2/{\rm d.o.f.}$ values for the fits using the RAR, taken from Li et al.~\cite{Li:2018rnd}. Interestingly, the RAR fits are comparable to the NFW fits, but much worse than SIDM and DC14 ones. We note that there are differences in the way that the different models were fit to the data. Li et al. fitted the whole SPARC sample with the relation, i.e., $175$ galaxies; while the SIDM and DC14 (NFW) fits included $149$ and $147$ galaxies, respectively. The extra galaxies in Li et. al. are those with small inclinations (less than 30 degrees) or galaxies with fewer than five data points or quality flag of 3 (with 1 and 2 defined as being ``good" in the SPARC sample). In addition, the priors on mass-to-light ratios are close, but not exactly the same. The SIDM and DC14 fits have two halo parameters for each galaxy, which is two more degrees of freedom in the fit compared to the RAR relation.  On the other hand, Li et al. marginalized galaxy distance and disk inclination to optimize the fits, but the other fits did not allow this additional freedom. 

The $\chi^2$ distribution of the fits indicate that the RAR does not provide a good explanation of the diverse rotation curves, which is explicit in the individual fits shown in Ren et al.~\cite{Ren:2018jpt}. For a more detailed comparison, we have revisited the RAR fits in~\cite{Ren:2018jpt} and compared the Bayesian Information Criterion (BIC) for the RAR and SIDM fits for $149$ galaxies. The BIC penalizes models with larger number of fit parameters ($k$) with an additive factor of $k*\ln(n)$, where $n$ is the number of data points for each galaxy. Despite this, we find that only a couple of galaxies strongly prefer the RAR fit over SIDM ($\Delta\rm BIC \geq 6$), demonstrating the importance of the scatter inherent in the SIDM fits in explaining the diversity. This is also evident from Fig. S1 in the supplemental material of Ren et. al., which clearly shows that the rotation curve shapes deviate systematically from the RAR in the inner regions for radii less than $2R_{\rm d}$. 

To further demonstrate this point, we use a more restricted SIDM model with only one-free parameter, the disk stellar mass-to-light ratio, $\Upsilon_{\star}$, as in the case of the RAR fits. We do so by {\em artificially} turning off the sources of scatter from the concentration-mass relation and the stellar-halo mass relation, i.e., we assume a median relation for them, without including their associated scatters. For an individual galaxy, a given value of $\Upsilon_{\star}$ determines the total stellar mass, from which we can infer the corresponding halo mass via the $M_{\star}\textup{--}{M}_{\rm vir}$ relation~\cite{2010ApJ...717..379B} and the scale radius through the concentration-mass relation~\cite{Dutton:2014xda} for the halo. We emphasize that turning off scatter is unphysical but it provides a convenient avenue to discussing the SIDM predictions in the RAR plane. In addition, this model has not been tuned to the rotation curves in any way -- the relations setting the halo parameters are simply taken from the literature. In the RAR fits, we take $g_\dagger=1.2\times10^{-10}~{\rm m/s^2}$ and use the form proposed in~\cite{McGaugh:2016leg}. To make the comparison as simple as possible, we fit $118$ SPARC galaxies that do not have a significant bulge and pass Katz et al.'s selection criteria~\cite{Katz:2016hyb} (see the discussion in Sec. IV). 

\begin{figure}[t]
\centering

\includegraphics[scale=0.53]{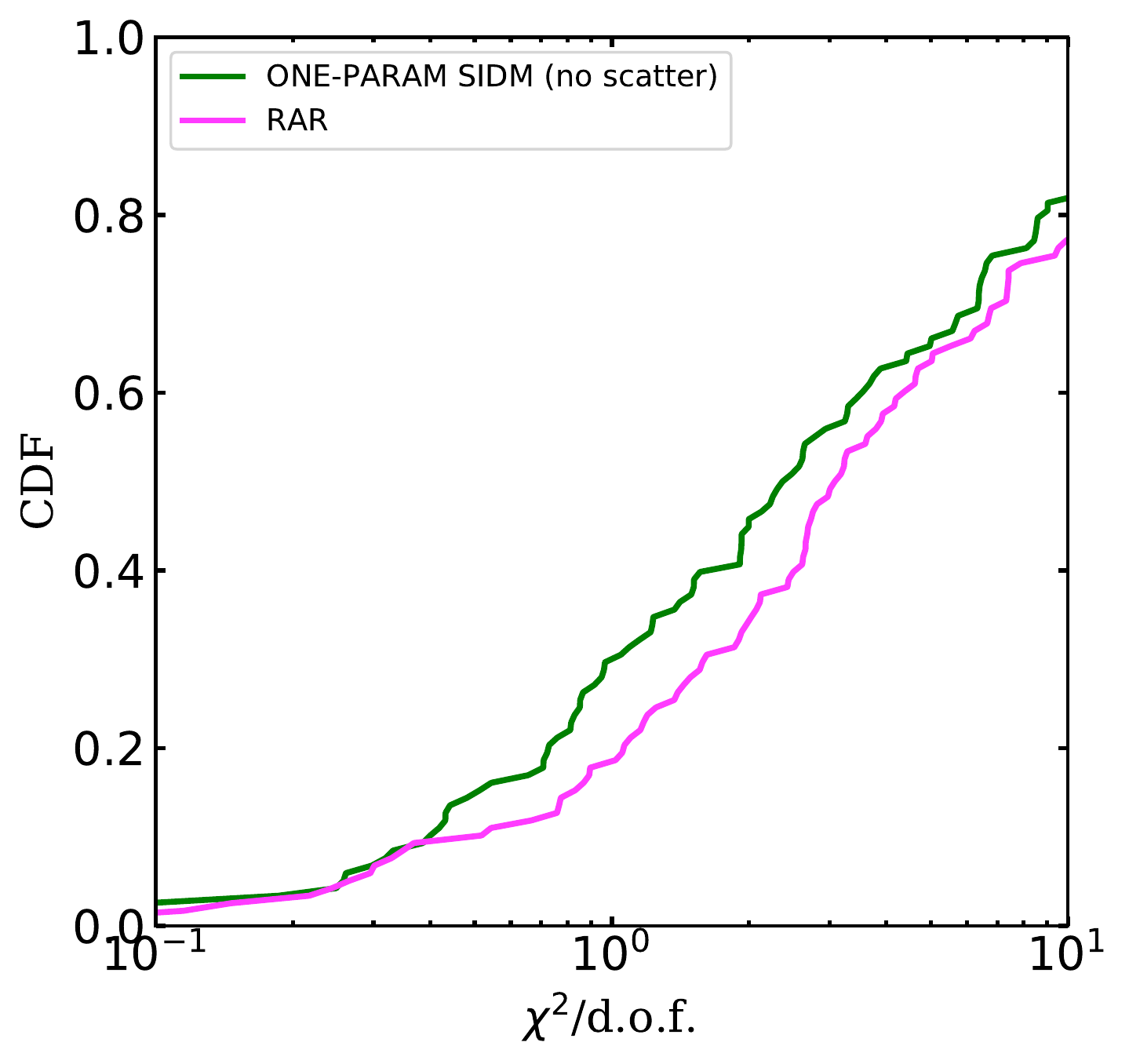} 
\includegraphics[scale=0.5]{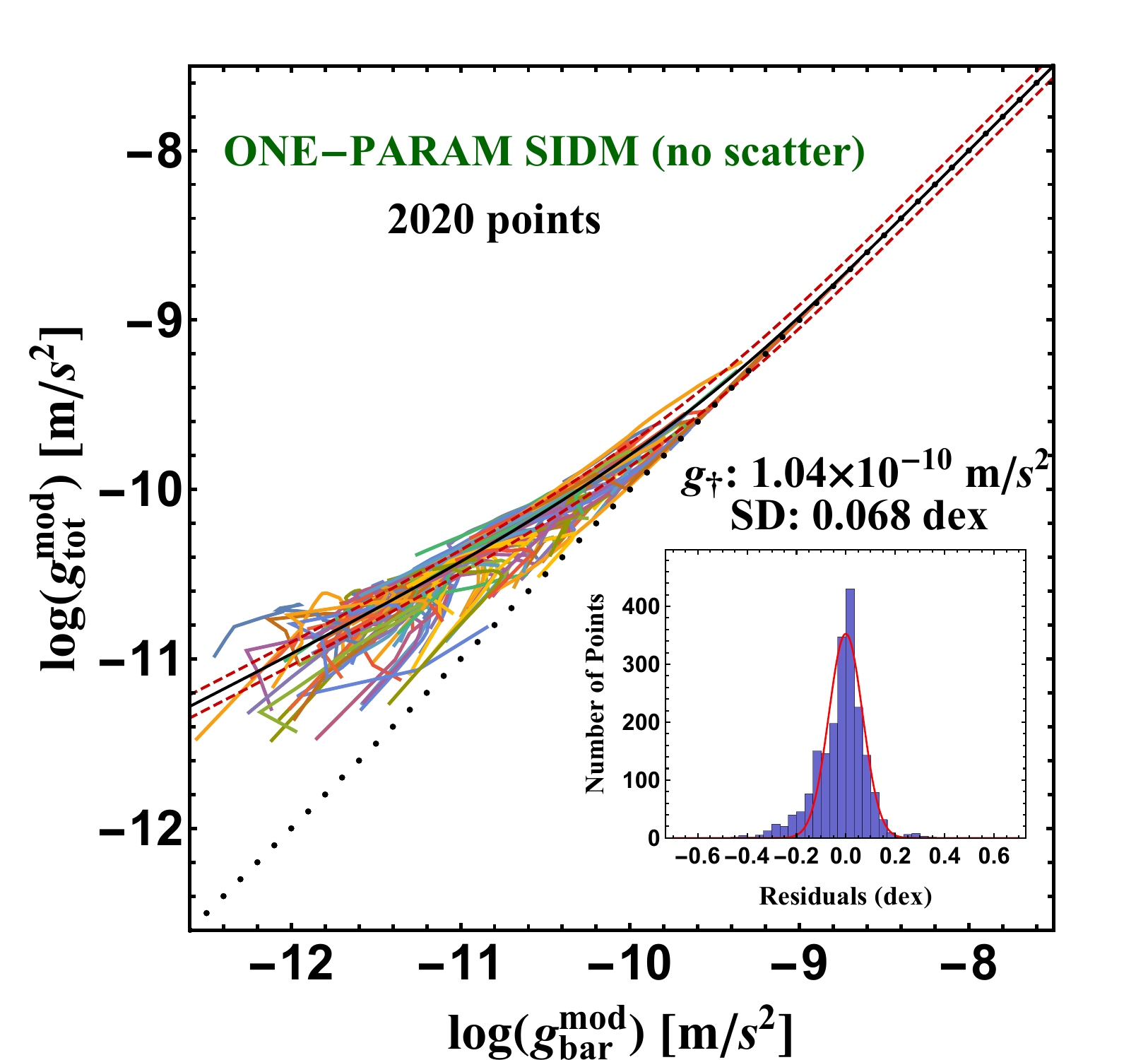} 

\caption{{\it Left:} The cumulative distribution functions of $\chi^2/{\rm d.o.f.}$ values for the maximum posterior rotation curve fits of the {\it one-parameter} SIDM model and the RAR to $118$ bulgeless galaxies in the SPARC sample.  {\it Right:} the total radial acceleration vs the baryonic acceleration, predicted in the one-parameter SIDM fits. The black solid line is the best to the RAR; the two red dashed curves correspond to the $1\sigma$ deviation from this fit. Inset: Corresponding histograms of residuals after subtracting the fit function with the best-fit scale parameter $g_\dagger=1.04\times10^{-10}~{\rm m/s^2}$. }
\label{fig:acc}
\end{figure}

Fig.~\ref{fig:acc} (left) shows the cumulative distributions of $\chi^2/{\rm d.o.f}$ values for the {\it one-parameter} SIDM and RAR fits. Since we do not allow scatters when imposing the priors, the one-parameter SIDM fits are worse than the corresponding fits shown in Fig.~\ref{fig:cdf}. However, the one-parameter SIDM fits are still better than the RAR fits, and we have checked that their $\Upsilon_\star$ distributions are similar, peaked around $0.5~{M_\odot/L_\odot}$. This could be because of the mild dependence of the acceleration scale on halo mass, predicted in hierarchical structure formation model, plays an important role in explaining the data, consistent with our earlier findings~\cite{Ren:2018jpt}, see also~\cite{2018NatAs...2..668R,Dutton:2019gor}. 

Furthermore, the one-parameter SIDM fits do not lead to a single line in the radial acceleration plane, as shown in Fig.~\ref{fig:acc} (right). This is because the predicted $g^{\rm mod}_{\rm tot}$--$g^{\rm mod}_{\rm bar}$ curve is {\em not} parallel to the RAR for points inside the stellar disk \cite{Ren:2018jpt}. This leads to a spread in $g^{\rm mod}_{\rm tot}$ that increases with decreasing $g^{\rm mod}_{\rm bar}$, even though the model has only one free parameter (stellar mass-to-light ratio) for each galaxy. When we fit a RAR to the predicted $g^{\rm mod}_{\rm tot}$ and $g^{\rm mod}_{\rm bar}$  following the procedure in Ref.~\cite{McGaugh:2016leg}, we find the best-fit characteristic acceleration scale is $g_\dagger=1.04\times10^{-10}~{\rm m/s^2}$ and the standard deviation of the corresponding residuals is $0.068$ dex. It is smaller than the one found in Ren et al.~\cite{Ren:2018jpt}, $0.10$ dex. This is not surprising because the latter takes into account the scatter in the $r_{\rm max}\textup{--}V_{\rm max}$ relation in the fits, which is important in explaining the full range of the diversity of the rotation curves. This is explicit proof that the correlations exhibited by the spiral galaxies do not imply that there has to be an underlying theory that predicts a one-to-one correlation between the total and and baryonic accelerations.

We conclude that the RAR obtained in~\cite{McGaugh:2016leg} is an average over all galaxies and getting models that lie along the RAR does not imply that the rotation curves are well fit. The hierarchical structure formation model predicts that there exists a characteristic acceleration scale  for each halo that depends mildly on the halo mass~\cite{vandenBosch:1999dz,Kaplinghat:2001me,Lin:2015fza,Navarro:2016bfs,Keller:2016gmw,Ren:2018jpt,Dutton:2019gor}. It also predicts the total rotation curve is determined by both halo and baryon contributions, and their relative contributions can vary significantly from low to high surface brightness galaxies. Both the concentration-mass relation and the stellar-to-halo mass relation play key roles in setting the radial acceleration relation in the averaged sense as described above. Both the SIDM and CDM models have these features built in, but the modification of inner dark matter distributions due to thermalization makes SIDM more consistent with observations.

\section{Conclusions}

In this work, we have compared SIDM and CDM explanations to the core vs cusp problem based on fits to the rotation curves of the SPARC samples, obtained from the literature. The SIDM model predicts cored dark matter density profiles in low surface brightness galaxies and cuspy density profiles in high surface brightness galaxies, and agrees best with observations. We showed explicitly that the logarithmic slope of the dark matter density profile at $1.5\%$ of the virial radius inferred from the SIDM fits is correlated with the stellar surface density. The CDM simulations with strong baryonic feedback can produce dark matter density cores on dwarf scales, providing a much better fit to the data than the NFW model. However, these models fail to reproduce galaxies with high stellar and dark matter densities. On the other hand, models with weak feedback are unable to produce galaxies with cored density profiles. We conclude that current feedback models need to be modified, and more work on regulating star formation in the context of both CDM and SIDM models is required. 

We provided a statistical comparison of fits with SIDM and CDM models, as well as the RAR, and found that the SIDM model is the best description of the rotation curves. We traced this to the fact that the SIDM model predicts both cored and cuspy dark matter density profiles, depending on stellar surface brightness. We constructed a one-parameter SIDM model by assuming the median concentration-mass and the stellar-to-halo mass relations with no scatter. Even in this case, the SIDM model does better than the RAR, which is only recovered in an averaged sense. These results demonstrate that thermalization of the inner halos due to dark matter self-interactions is currently the preferred solution to the core vs cusp, aka diversity, problem. 

\acknowledgments
We thank Isabel M. Santos-Santos for providing us the simulation data used in Fig. 1 and Federico Lelli for useful discussion. This work was supported by the National Science Foundation Grant No.~PHY-191505 (MK), the U. S. Department of Energy under Grant No.~DE-SC0008541 (HBY) and UCR Academic Senate Regents Faculty Development Award (HBY).

\bibliography{sidm}

\end{document}